\PassOptionsToPackage{x11names,table,dvipsnames}{xcolor}
\documentclass[acmtog,balance=false,authorversion]{acmart}

\usepackage{subcaption}
\usepackage{bm}
\usepackage{wrapfig}
\usepackage[ruled]{algorithm2e} %
\usepackage{color}

\SetAlFnt{\small}
\SetAlCapFnt{\small}
\SetAlCapNameFnt{\small}
\SetAlCapHSkip{0pt}

\setcopyright{acmlicensed}
\acmJournal{TOG}
\acmYear{2025} \acmVolume{1} \acmNumber{1} \acmArticle{1} \acmMonth{1}\acmDOI{10.1145/3767323}

\definecolor{redcolor}{rgb}{0.8,0,0}
\definecolor{bluecolor}{rgb}{0,0,0.8}
\definecolor{greencolor}{rgb}{0.0,0.5,0.0}

\allowdisplaybreaks

\begin{document}
\title{Local Surface Parameterizations via Smoothed Geodesic Splines}

\begin{teaserfigure}
  \includegraphics[width=\textwidth]{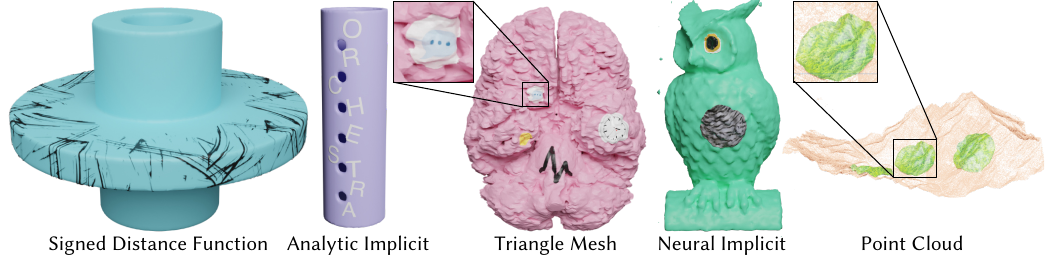}
  \caption{The local parameterizations produced by our method provide low-distortion maps on a variety of different types of geometry. We can place scratches on a well-used children's toy represented by a signed distance function (SDF); stick letters onto a flute composed with CSG operations, which as a result does not behave like an SDF; place pictorial labels on a triangle mesh of a brain reconstructed from an MRI scan (zoom in to see more details); fill in one eye and some plumage of an owl represented by a neural implicit function; and even add some forest decals onto a terrain point cloud with over 29 million points acquired from LiDAR data.}
  \label{fig:teaser}
\end{teaserfigure}

\author{Abhishek Madan}
\affiliation{%
  \institution{University of Toronto}
  \city{Toronto}
  \country{Canada}
}
\email{amadan@cs.toronto.edu}
\author{David I.W. Levin}
\affiliation{%
  \institution{University of Toronto and NVIDIA}
  \city{Toronto}
  \country{Canada}
}
\email{diwlevin@cs.toronto.edu}

\newcommand{\bx}{\mathbf{x}}
\newcommand{\by}{\mathbf{y}}
\newcommand{\bp}{\mathbf{p}}
\newcommand{\bv}{\mathbf{v}}
\newcommand{\bq}{\mathbf{q}}
\newcommand{\bn}{\mathbf{n}}
\newcommand{\bt}{\mathbf{t}}
\newcommand{\ba}{\mathbf{a}}
\newcommand{\bzero}{\mathbf{0}}

\newcommand{\T}{\mathcal{T}}
\newcommand{\M}{\mathcal{M}}
\newcommand{\mS}{\mathcal{S}}

\newcommand{\R}{\mathbb{R}}
\newcommand{\bV}{\bar{V}}
\newcommand{\bF}{\bar{F}}

\newcommand{\Lap}{\mathsf{L}}
\newcommand{\Id}{\mathsf{I}}

\newcommand{\bphi}{\bm{\phi}}
\newcommand{\blambda}{\bm{\lambda}}
\newcommand{\bpi}{\bm{\pi}}
\newcommand{\btheta}{\bm{\theta}}
\newcommand{\bTheta}{\bm{\Theta}}
\newcommand{\bPhi}{\bm{\Phi}}
\newcommand{\btau}{\bm{\tau}}

\newcommand{\Tspace}[1]{\T_{#1}\M}
\newcommand{\Tplane}[1]{T_{#1}\M}

\begin{abstract}
We present a general method for computing local parameterizations rooted at a point on a surface, where the surface is described only through a signed implicit function and a corresponding projection function.
Using a two-stage process, we compute several points radially emanating from the map origin, and interpolate between them with a spline surface.
The narrow interface of our method allows it to support several kinds of geometry such as signed distance functions, general analytic implicit functions, triangle meshes, neural implicits, and point clouds.
We demonstrate the high quality of our generated parameterizations on a variety of examples, and show applications in local texturing and surface curve drawing.
\end{abstract}



\maketitle

\section{Introduction}

Local parameterizations are particularly useful geometric tools mapping a planar coordinate system to a local, manifold region around a specified origin, and enable graphics tasks like placing decals and small texture patches on surfaces (Fig.~\ref{fig:teaser}).
Local parameterizations are particularly well suited for complex geometry, trading a global parameterization problem that must deal with topologically mandated difficulties like singularity placements and seams~\cite{Levy2002,Mullen2008,Soliman2018} for a local problem with looser preconditions.

However, most of the approaches developed for computing local parameterizations only work on meshes, or more generally, graphs. 
This allows them to leverage existing tools such as discrete differential operators~\cite{Sharp2019,Herholz2019} or Dijkstra's algorithm~\cite{Schmidt2006,Schmidt2013,Melvaer2012},
but ties output quality to sampling or mesh quality. 
The sources of geometric representations in modern applications are quite varied (from large artist-generated meshes, to LiDAR-scanned point clouds, to hand-crafted implicits, and even neural network-encoded implicit functions~\cite{Takikawa2021,Muller2022})
and for many of these, it can be difficult to extract the high-quality samples and topological information required to apply previous methods.
Meshes, for instance, may have low-quality triangles that are unsuitable for finite element-based computation; point clouds lack explicit topology altogether and may contain noise from the acquisition procedure; and implicit functions lack both explicit geometry and topology, with possibly noisy isosurfaces in the case of neural implicits.
Many methods exist to address these problems, from remeshing~\cite{Botsch2004}, to surface reconstruction~\cite{Lorensen1987,Ju2002,Chen2021,Chen2022,Sellan2023}, and even sampling the surface~\cite{Witkin1994,DeGoes2012}, but these methods are quite disparate, and as such there is no unified way to generate local parameterizations for all these types of representations.

We propose a technique to avoid such preprocessing by only requiring a very narrow interface from the input geometry: a signed implicit function, along with a projection function derived from this implicit function.
It is typically straightforward to compute this implicit representation: for example, obtaining an implicit function from meshes is simply a matter of computing the unsigned minimum distance and signing it with the winding number~\cite{Jacobson2013,Barill2018}.
Our algorithm proceeds in two stages: first tracing geodesic-like paths from a prescribed origin, and then fusing them into a continuous local map using a spline surface.
Tracing geodesics along surfaces via projection is a simple yet surprisingly delicate procedure (Fig.~\ref{fig:projection}), so we also present a curvature-sensitive substepping procedure that avoids path-splitting steps in high-curvature regions, as well as an inter-curve smoothing method that greatly reduces map artifacts created by large variations in curvature across the surface.
To our knowledge, our method is the first to develop a radial tracing method for local parameterizations that explicitly addresses both of these issues.
By exclusively relying on pointwise implicit function queries rather than global surface operators, we obtain an output-sensitive method (i.e., where computational work scales in proportion to the size of the map) that produces local parameterizations meeting or exceeding the quality of previous methods, particularly on surfaces other than high-quality triangle meshes.
We demonstrate the efficacy of our approach on a wide variety of geometric representations and regions on these geometries.

\section{Related Work}

\begin{figure}
  \includegraphics[width=\columnwidth]{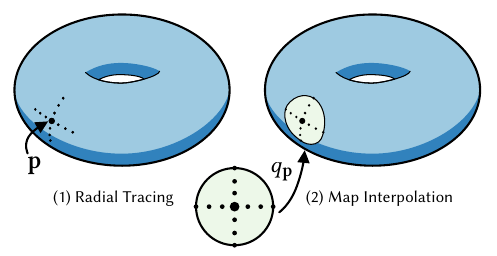}
  \caption{Our method consists of two main components: tracing out points along radial curves from an origin $\bp$ (left), and interpolating those points to form a continuous map $q_\bp$ (right).}
  \label{fig:overview}
\end{figure}

\begin{figure}
  \includegraphics[width=\columnwidth]{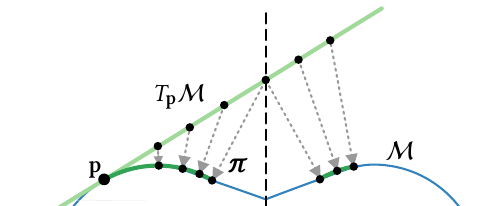}
  \caption{Farther away from the origin $\bp$ on $\Tplane{\bp}$, the (unsmoothed) projection image of a tangent line onto $\M$ can shrink significantly, and even split into disconnected pieces on $\M$ past the medial axis (vertical dashed line).}
  \label{fig:projection}
\end{figure}

\textit{Computing geodesics.}
Several methods have been developed to compute geodesics on meshes, both exact and approximate~\cite{Polthier2006,Surazhsky2005,Kimmel1998,Crane2017}.
This problem comes in many flavours, from tracing out geodesic paths, to finding globally shortest geodesics between two given points, and even to finding all geodesics from one point to the rest of a set of points --- see \citet{Crane2020} for a more comprehensive survey.
Many of these methods can operate in a purely intrinsic setting due to them having access to an underlying mesh, which allows them to avoid an extrinsic projection-like procedure, but as such they are limited in the kinds of representations they can operate on.
On the other hand, our work removes the underlying mesh assumption and works on a broader set of representations, though we note that we focus only on the geodesic tracing problem to build our parameterizations, rather than the full set of geodesic problems that have been studied.
Some recent work also explores regularized geodesic distances~\cite{Edelstein2023}, but this is primarily restricted to meshes as well.
While there has been some prior work on computing geodesics on implicit surfaces which also leverages a projection operator derived from the implicit function~\cite{Pedersen1995}, it requires an underlying blue noise global sampling of the surface to warm start the procedure, which our method avoids.
Projection can also behave poorly in high-curvature regions (Fig.~\ref{fig:projection}), which is why many previous methods prefer intrinsic mesh-based operations; however, we utilize a substepping procedure that preemptively avoids difficult projection scenarios, making it a feasible building block for our parameterization.

\textit{Parallel transport.}
A concept closely related to geodesics is parallel transport, an operation which provides a way to connect tangent spaces on a manifold.
For example, tracing out geodesics is equivalent to parallel transporting the geodesic's tangent vector across the surface, while simultaneously following along that tangent.
One early method for discretizing parallel transport is Schild's ladder, which approximates parallel transport by drawing geodesic parallelograms on the surface.
Although it has found some use in image analysis~\cite{Lorenzi2014}, it requires a method to find geodesics and as such does not help to compute geodesics on its own.
There exist multiple frameworks for parallel transport on meshes which operate on different subsimplices, such as faces and the dual edges connecting them~\cite{Crane2010}, and vertices~\cite{Knoppel2013}.
A practical mesh-free way to compute parallel transport involves computing the smallest rotation to align nearby tangent planes~\cite{Schmidt2006}, which is the approach we use as well.

\textit{Local parameterizations.}
Local parameterizations are also a well-explored topic in graphics.
\citet{Pedersen1995} presents an interactive system for drawing paramaterization boundaries for rectangular patches, but this approach can be difficult to implement and use robustly, and also requires global surface samples.
As such, many interactive local parameterization methods compute maps about a fixed origin instead, and let the algorithm expand the map on the surface rather than having the user guess appropriate boundaries.
Local parameterization methods have also been developed using neural networks~\cite{Groueix2018,Williams2019,Srinivasan2023}, but like \citet{Pedersen1995}, these methods do not center about a specific point and also must be trained on surface samples (though it is worth noting that the aim of these neural methods is to produce an ensemble of local charts for a global chart, so they are less concerned with user-defined initial conditions).

\textit{Exponential maps.}
Another local parameterization approach is based on the exponential map, which maps tangent space points onto the surface around an origin $\bp$.
Discrete exponential maps are one example of such a method~\cite{Schmidt2006,Schmidt2013}; despite the method name, it computes the logarithmic map (i.e., the inverse of the exponential map), given a graph connecting the point to nearby surface samples.
As a result, the method is very sample-dependent, producing bad results in regions that are insufficiently sampled relative to their curvature.
Another Dijkstra-like method is by \citet{Melvaer2012}, which infers appropriate distance information by leveraging triangle mesh connectivity rather than a general graph.
More recently, the vector heat method~\cite{Sharp2019} computes logarithmic maps using a connection Laplacian operator on the surface to approximate parallel transport, but the result quality is also strongly tied to the underlying triangulation (even with the aid of intrinsic Delaunay remeshing).
Although the method only requires a way to compute the required operator matrix and thus is not strictly tied to triangle meshes, it can struggle to produce good results on poorly distributed point clouds, even when leveraging a high-quality point cloud Laplacian algorithm~\cite{Sharp2020}.
A similar method was concurrently proposed by \citet{Herholz2019}, which also uses a heat method to compute log maps, but uses angles between distance gradients to approximate geodesic polar angles.
Another drawback of PDE-based approaches is that they solve a global linear system to compute a local parameterization.
Like \citet{Schmidt2006} and \citet{Schmidt2013}, our method is output-sensitive, doing work proportional to the size and complexity of the region to be parameterized, independent of the rest of the surface. 
This is because our method generates the samples it needs on the fly, and can be evaluated to produce high-quality maps on even poor sample sets.
Local parameterizations based on exponential maps have also been used to develop a generalized convolution neural architecture for triangle meshes~\cite{Masci2015}, where geodesics are estimated by triangle unfolding, which (unlike our approach) limits it to triangle meshes by construction.

\section{Method}

The theoretical framework for our local parameterization technique is based on the exponential map.
Mathematically speaking, an exponential map $\exp_\bp$ is a function rooted at a point $\bp$ on a manifold $\M$, which maps tangent vectors $\bt$ from $\bp$'s tangent space $\Tspace{\bp}$ (represented by vectors in $\R^2$) and has the following properties:
\begin{itemize}
  \item $\exp_\bp(\bzero) = \bp$ (i.e., the zero tangent vector maps to the origin $\bp$); and
  \item $\exp_\bp(\bt) = \gamma(\|\bt\|)$, where $\gamma$ is an (arc-length parameterized) geodesic described by the initial conditions $\gamma(0) = \bp$ and $\gamma^\prime(0) = \bt/\|\bt\|$.
\end{itemize}

Essentially, exponential maps describe how to trace out geodesics emanting from $\bp$ onto $\M$.
As such, they exhibit low distortion near the origin and are thus a very useful construction to build upon for local parameterizations.
This radial tracing interpretation also suggests a two-step process for generating a local parameterization around $\bp$: (1) trace out several geodesic-like curves from $\bp$ (Sections~\ref{sec:trace}--\ref{sec:smoothing}), and (2) interpolate these curves radially to produce a continuous map (Section~\ref{sec:interp}).
Fig.~\ref{fig:overview} contains a visual overview of the algorithm.

Key to the generality of our approach is the limited interface it requires from input geometry.
We assume that we are only given an implicit function $f(\bx)$ which contains the surface of interest in its zero isosurface $\M$.
From $f$, we can obtain a gradient $\nabla f(\bx)$, a normal $\bn(\bx) = \frac{\nabla f(\bx)}{\| f(\bx) \|}$ (defined only on $\M$), and a projection operator $\bpi(\bx)$ onto $\M$.
We define $\bpi$ based on the general projection procedure described in~\citet{Atzmon2019}, which generalizes signed distance projection through an iterative root-finding procedure $\overline{\bx}_{i+1} = \overline{\bx}_i - f(\bx) \frac{\nabla f(\bx)}{\| \nabla f(\bx) \|^2}$, $\overline{\bx}_0 = \bx$.
It is worth noting that, in order to have a well-defined $\bn$ (and hence well-defined tangent planes), 0 must not be a local extremum of $f$; in particular, unsigned distance functions do not work unless they are offset to have a different zero isosurface.
Also, unless otherwise stated, we use a smoothed gradient $\widetilde{\nabla f}(\bx) = \left( \int_{B_{\bx,\epsilon}} \nabla f(\by) d\by \right) / | B_{\bx,\epsilon} |$ instead of $\nabla f(\bx)$ to define normals, tangent planes, and projections, where $B_{\bx,\epsilon}$ is the ball of radius $\epsilon$ centered at $\bx$ and $|B_{\bx,\epsilon}|$ is its volume, to remove $C^1$ discontinuities in $f$.
This essentially turns the medial axis into regions of rapidly changing gradients, which becomes useful for detecing large accumulated curvature (Section~\ref{sec:substep}).

\subsection{Radial Tracing}\label{sec:trace}

We wish to find a local parameterization about a point $\bp \in \M$, which we denote by $q_\bp \colon \Tspace{\bp} \to \M$ (we do not write it as $\exp_\bp$ since, as we will describe later, we do not always want an exact exponential map).
To start, we set $q_\bp(\bzero) = \bp$, and then we numerically integrate $m$ equally spaced geodesic-like radial curves radiating from $\bp$, with initial tangent directions defined extrinsically in $\R^3$.
Using $\Tplane{\bp}$ to represent the tangent plane at $\bp$ to $\M$ (i.e., the embedding of $\Tspace{\bp}$ in $\R^3$), these tangent directions are denoted by $\bt_{i,0} \in \Tplane{\bp}$, $0 \le i \le m-1$.
We perform a single integration step in direction $i$ by moving along $\bt_{i,0}$ for a prescribed step size $h$, and projecting to $\M$:
\begin{equation}
  \label{eq:step0}
  \bq_{i,1} = \bpi(\bp + h\bt_{i,0}).
\end{equation}
For small $h$, this is a good approximation to integrating a short geodesic from $\bp$ along the surface of $\M$, where the exact geodesic path's preimage through $\bpi$ on $\Tplane{\bp}$ would be a curved path. %

To continue integrating the radial curves for more steps, we now need to parallel transport the $\bt_{i,0}$'s to their corresponding $\bq_{i,1}$ points, or in other words, transfer them from $\Tplane{\bp}$ to the tangent plane at $\bq_{i,1}$, $\Tplane{\bq_{i,1}}$.
We can do this by computing the smallest rotation that transforms $\Tplane{\bp}$ into $\Tplane{\bq_{i,1}}$, denoted $R_{i,0}$, and applying it to $\bt_{i,0}$: $\bt_{i,1} = R_{i,0} \bt_{i,0}$.
More concretely, $R_{i,0} = \exp [ \alpha_{i,0} \hat{\ba}_{i,0} ]$, where $\alpha_{i,0} = \arccos( \bn(\bp) \cdot \bn(\bq_{i,1}) )$, $\hat{\ba}_{i,0} = \frac{\bn(\bp) \times \bn(\bq_{i,1})}{\| \bn(\bp) \times \bn(\bq_{i,1}) \|}$, and $\exp[ \bv ]$ is the matrix exponential of the skew-symmetric cross product matrix of $\bv$.
According to Minding's theorem, $\bt_{i,1}$ is precisely the result of parallel transporting $\bt_{i,0}$ to $\bq_{i,1}$ for constant curvature surfaces, and therefore it is a good approximation for nearby points whose local neighbourhoods are similar~\cite{Schmidt2006}.
Using the $\bt_{i,1}$'s, we can repeat the integration step described earlier, but where $\bq_{i,1}$ replaces $\bp$ along each path $i$.
For a general step $j+1$, we therefore have:
\begin{equation}
  \label{eq:step}
  \bq_{i,j+1} = \bpi(\bq_{i,j} + h\bt_{i,j}).
\end{equation}
To more explicitly show the correspondence between Eq.~\ref{eq:step0} and Eq.~\ref{eq:step}, we can denote $\bp$ by $\bq_{i,0}$.
We repeat for $n$ iterations until we have the set of traced points $Q = \{ \bq_{i,j} \mid 0 \le i \le m-1, 0 \le j \le n \}$, which can be arranged into the set of radial curves $\Gamma = \{ \gamma_0, \gamma_1, \ldots, \gamma_{m-1} \}$ where $\gamma_i = [ \bq_{i,0}, \ldots, \bq_{i,n} ]$, as well as the set of isolines $\Psi = \{ \psi_0, \psi_1, \ldots, \psi_n \}$ where $\psi_j = [ \bq_{0,j}, \bq_{1,j}, \ldots, \bq_{m-1,j} ]$ (note that $\psi_0$ is simply $\bp$).
We treat $\gamma_i$ and $\psi_j$ as both point sequences and curves connecting the points, depending on context.

Pseudocode describing this step, along with the improvements described in Sections~\ref{sec:substep} and~\ref{sec:smoothing}, is given in Alg.~\ref{alg:trace}.

\begin{algorithm}
  \caption{Radial tracing \textsc{radialTrace}}\label{alg:trace}
  \SetAlgoLined
  \LinesNumbered
  \SetKwInOut{Input}{Inputs}
  \SetKwInOut{Output}{Outputs}
  \Input{Implicit function $f$, \# of radial curves $m$, \# of steps $n$, step size $h$}
  \Output{Point samples $Q$}
  \BlankLine
  $Q \leftarrow \{ \bp \}$\;
  \For{$j \leftarrow 0, \ldots, n-1$}{
    \For{$i \leftarrow 0, \ldots, m-1$}{
      \tcp{Section~3.2}
      $\tilde{h} \leftarrow h$\;
      $\tilde{\bq}_{i,j} \leftarrow \bq_{i,j}$\;
      $\tilde{\bt}_{i,j} \leftarrow \bt_{i,j}$\;
      \While{$\tilde{h} \ge 10^{-6}$}{
        $\ell \leftarrow$ solve Eq.~\ref{eq:alignment} in $\left[ 0,\tilde{h} \right]$\;
        $\tilde{\btau}_{i,j} \leftarrow \tilde{\bq}_{i,j} + \ell\tilde{\bt}_{i,j}$\;
        $\tilde{h} \leftarrow \tilde{h} - \| \bpi(\tilde{\btau}_{i,j}) - \tilde{\bq}_{i,j} \|$\;
        Transport $\tilde{\bt}_{i,j}$ from $\tilde{\bq}_{i,j}$ to $\bpi(\tilde{\btau}_{i,j})$\;
        $\tilde{\bq}_{i,j} \leftarrow \bpi(\tilde{\btau}_{i,j})$\;
      }
    }
    \tcp{Section~3.3}
    $\bTheta_{j+1} \leftarrow$ solve Eq.~5\;
    \For{$i \leftarrow 0, \ldots, m-1$}{
      $\bq_{i,j+1} \leftarrow \tilde{\bq}_{i,j}$\;
      $\bt_{i,j+1} \leftarrow$ rotate $\tilde{\bt}_{i,j}$ about $\bn(\bq_{i,j+1})$ by $\theta_{i,j+1}$\;
      $Q \leftarrow Q \cup \{ \bq_{i,j+1} \}$\;
    }
  }
  Return $Q$\;
\end{algorithm}

\subsection{Substepping}\label{sec:substep}

Although $h$ is small, projecting a straight line from a general $\Tplane{\bq_{i,j}}$ onto $\M$ can still behave poorly far away from the plane origin: its projection image can shrink significantly to almost zero length in regions of very high curvature, and can even split near the medial axis (Fig.~\ref{fig:projection}).
Since $\bn$ is a smoothed normal, it already prevents image splitting, but can instead lead to deformed projection images, which significantly degrade the quality of the final parameterization.
Therefore, we sometimes need to take substeps smaller than $h$ to ensure that each full step has a geodesic length of roughly $h$,
which is a technique commonly used when numerically integrating differential equations in order to obtain higher accuracy~\cite{dahlquist2008numerical}.

To achieve a uniform step size across all directions, we find a point along each straight line $\btau_{i,j}(\ell) = \bq_{i,j} + \ell\bt_{i,j}$ which satisfies the following \textit{projected alignment equation} for each $i$:
\begin{equation}
  \label{eq:alignment}
  \begin{aligned}
    \bn(\bq_{i,j}) \cdot \bn(\bpi(\btau_{i,j}(\ell))) &= s \\
    \text{s.t. } \ell & \in [0,h],
  \end{aligned}
\end{equation}
where $s$ is a prescribed alignment cosine (we used $s = \cos \frac{\pi}{4} = \frac{1}{\sqrt{2}}$ in all our experiments).
This equation can be solved using a general non-linear root finding procedure such as the bisection method, and if no solution is found, we return the upper bound of the interval, $h$.
Essentially, we want to find the point along $\btau_{i,j}$ where its projected normal significantly deviates from the plane normal, which indicates a large accumulation of curvature and/or having approached a normal discontinuity on $\M$.
To further ensure that we are taking full steps of approximate geodesic length $h$, we estimate our projected step size for the substep as $\| \bpi(\btau_{i,j}(\ell)) - \bq_{i,j} \|$, subtract it from $h$ to get a new max step size, and repeat the procedure until the max step size is sufficiently small (see Alg.~\ref{alg:trace}).

\subsection{Holonomy Smoothing}\label{sec:smoothing}

\begin{figure}
  \includegraphics[width=\columnwidth]{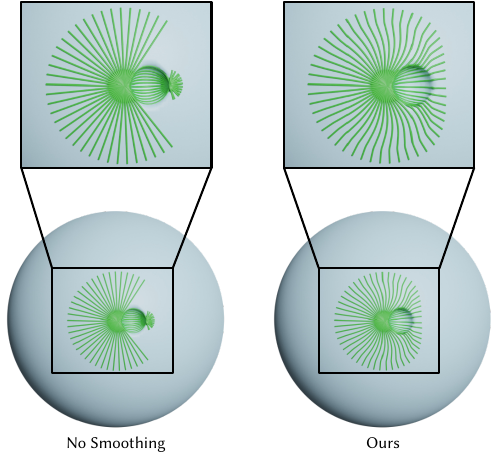}
  \caption{Without holonomy smoothing (left), the traced paths can drastically separate and intersect with each other away from the origin after they encounter a small pothole. Smoothing out the wedge holonomies (right) causes the paths to more closely emulate their trajectories on a similar surface with constant curvature (in this case, a sphere without a pothole).}
  \label{fig:holonomy_comp}
\end{figure}

Aside from high curvature along a path on the surface, another issue is the variation in curvature between nearby geodesics.
The relationship between nearby geodesics is captured by the \textit{Jacobi equation}~\cite{docarmo1992riemannian,pottmann2010geodesic}, where the acceleration of the separation between geodesics is proportional to the negative Gaussian curvature of the surface.
As a result, geodesics accelerate towards each other on surfaces with positive Gaussian curvature, accelerate away from each other on surfaces with negative Gaussian curvature, and retain their separation velocity on surfaces with zero Gaussian curvature (Fig.~\ref{fig:holonomy_comp}, left).
When acceleration rates vary significantly between traced geodesics, downstream interpolation becomes more challenging, because some regions of the map exhibit significant stretching and poorly represent the underlying surface, while others experience foldovers and lose injectivity (Fig.~\ref{fig:holonomy_regularization}, right). 
These issues are prominent on surfaces with high frequency details, or even ``noisy'' surfaces like those from neural implicits.

Since the underlying issue is variance in acceleration, we must simultaneously correct all traced radial curves to give them more uniform acceleration.
One possible approach is to adjust the traced points directly, but this would require a difficult constrained optimization over surface point positions.
Instead, we use the following simplifications: (1) we perform an optimization after each (full) integration step $j$, and (2) we optimize over rotations in the tangent plane that, at a high level, ``smooth out'' the traced paths over the integration front $\psi_j$.
Optimizing rotations in this way allows the projected steps do most of the work while we nudge the paths in better directions as we progress.

\setlength{\columnsep}{0.7em}
\setlength{\intextsep}{0em}
\begin{wrapfigure}{r}{0.52\columnwidth}
  \centering
  \includegraphics[width=0.52\columnwidth]{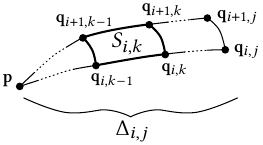}
\end{wrapfigure}
We model our optimization on the case of constant acceleration, to encourage the curves to spread out as uniformly as possible.
The surfaces satisfying this ideal property are those with constant Gaussian curvature, so the optimization procedure should encourage the curves to behave as if they were on a constant curvature surface.
On such surfaces, we expect the total Gaussian curvature in the wedges between curves to be equal, where wedges are defined as $\Delta_{i,j} = [ \bp, \ldots, \bq_{i,j}, \bq_{i+1,j}, \ldots, \bp ]$ (see inset) where each point is connected by a geodesic to the next point in the sequence and $i$ indices are modulo $m$.
Rather than directly integrating the curvature within this loop, which is difficult to do on an implicitly defined surface, we will instead use \textit{holonomy}, which is defined as the amount of rotation induced by parallel transport along a closed loop on the surface, and equivalently the total Gaussian curvature contained in that loop on the surface.
This way, we can convert an integration problem into a parallel transport problem, which we are well-equipped to solve.
Since we want to find in-plane rotations, we represent the rotations by angles $\theta_{i,j}$ in each $\Tspace{\bq_{i,j}}$ and solve for these angles.
(More precisely, we are now measuring the holonomy of a slightly modified $\Delta_{i,j}$ where infinitesimally small geodesics are extended from each $\bq_{i,j}$ after rotating each geodesic tangent vector by $\theta_{i,j}$, though with a slight abuse of notation we will refer to both the infinitesimally extended wedge and the original wedge by $\Delta_{i,j}$, as we are exclusively concerned with the infinitesimally extended version.)
Denoting the holonomy in a wedge by $R(\Delta_{i,j})$, we smooth out the wedge holonomies by minimizing a Dirichlet-like energy to obtain rotations $\bTheta_j = [ \theta_{0,j}, \theta_{1,j}, \ldots, \theta_{m-1,j} ]$:
\begin{equation}
  \label{eq:holonomy_opt}
  \bTheta^*_j = \min_{\bTheta_j} \sum_i R(\Delta_{i,j})^2 + \frac{1}{\kappa} \| \bTheta_j \|^2.
\end{equation}

\begin{figure}
  \includegraphics[width=\columnwidth]{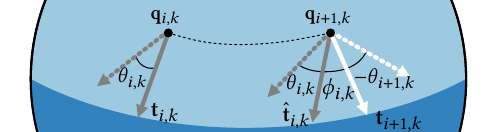}
  \caption{An illustration of measuring the change in angle as $\bt_{i,k}$ is transported from $\Tspace{\bq_{i,k}}$ to $\Tspace{\bq_{i+1,k}}$ (represented as $\hat{\bt}_{i,k}$ in the latter space). Solid arrows represent tangent vectors before holonomy smoothing, and dashed arrows represent tangent vectors after holonomy smoothing. All angles are measured clockwise.}
  \label{fig:holonomy}
\end{figure}

We measure total wedge holonomy by decomposing the wedge loop into strips $S_{i,k} = [ \bq_{i,k-1}, \bq_{i,k}, \bq_{i+1,k}, \bq_{i+1,k-1}, \bq_{i,k-1} ] $ (see inset), where each point is connected by a geodesic just like in the definition of $\Delta_{i,j}$, and then adding the holonomies of these strips.
Starting at $\bq_{i,k-1}$, our goal is to transport $\bt_{i,k-1}$ around $S_{i,k}$. and to simplify the result, we will set each $\bt_{p,q}$ to have an angle of 0 in its corresponding tangent space.
By construction, transporting along the geodesic from $\bq_{i,k-1}$ to $\bq_{i,k}$ does not change the angle, and the same is true from $\bq_{i+1,k}$ to $\bq_{i+1,k-1}$, so the main challenge is transporting from $\bq_{i,k}$ to $\bq_{i+1,k}$ ($\bq_{i+1,k-1}$ to $\bq_{i,k-1}$ is similar).
When the distance between $\bq_{i,k}$ and $\bq_{i+1,k}$ is small, we can approximately transport vectors from $\Tspace{\bq_{i,k}}$ to $\Tspace{\bq_{i+1,k}}$ using the smallest rotation that aligns $\Tplane{\bq_{i,k}}$ with $\Tplane{\bq_{i+1,k}}$, in the same manner by which we transported tangent vectors along short geodesic segments during tracing.
Then, we can find the change in angle (induced by a change of basis between tangent spaces) by measuring the angle between the transported $\bt_{i,k}$ (denoted $\hat{\bt}_{i,k}$) and $\bt_{i+1,k}$ in $\Tspace{\bq_{i,k}}$, which is just the signed angle from $\bt_{i+1,k}$ to $\hat{\bt}_{i,k}$ in $\R^3$.
When $\theta_{i,k} = \theta_{i+1,k} = 0$ (i.e., before smoothing), we denote this angle by $\phi_{i,k}$
(and all such angles at step $k$ by $\Phi_k$),
so in the general non-zero case, we have an angle of $\theta_{i,k} + \phi_{i,k} - \theta_{i+1,k}$ (Fig.~\ref{fig:holonomy}).
The change in angle from $\bq_{i+1,k-1}$ to $\bq_{i,k-1}$ has a similar expression but negated since we travel in the opposite direction with respect to the indices; thus we have
\begin{equation}
  \label{eq:strip_holonomy}
  R(S_{i,k}) = (\theta_{i,k} + \phi_{i,k} - \theta_{i+1,k}) - (\theta_{i,k-1} + \phi_{i,k-1} - \theta_{i+1,k-1}).
\end{equation}
Since we do not wish to rotate the initial tangent vectors $\bt_{i,0}$ and they are all in $\Tplane{\bp}$, we have $\theta_{i,0} = 0$ and $\phi_{i,0} = -\frac{2\pi}{m}$.

When we sum up the $R(S_{i,k})$, the $\theta_{i,k} + \phi_{i,k} - \theta_{i+1,k}$ terms cancel out when $1 \le k < j$, which gives
\begin{equation}
  \label{eq:holonomy}
  R(\Delta_{i,j}) = \sum_{k=1}^j R(S_{i,k}) = \theta_{i,j} + \phi_{i,j} - \theta_{i+1,j} + \frac{2\pi}{m}.
\end{equation}
Surprisingly, this expression does not directly depend on the angles obtained from previous steps, but in fact the change in angle terms $\Phi_j$ implicitly encode this information, because they are derived from the tangent directions that were obtained from the previous step's smoothing angles $\Theta_{j-1}$; in turn, $\Theta_{j-1}$ depends on the previous step's $\Phi_{j-1}$, which depend on $\Theta_{j-2}$, and so on.
Unrolling the dependency chain reveals that $\Phi_j$ contains information from all previous holonomy smoothing steps.

With the definition of $R(\Delta_{i,j})$, we can now see that Eq.~\ref{eq:holonomy_opt} resembles a 1D Dirichlet problem with periodic boundary conditions, which can be solved with a circulant tridiagonal linear system.
The linear system contains a null space and can be poorly conditioned, so we also include a regularization term $\frac{1}{\kappa} \| \bTheta_j \|^2$ to mitigate these issues, where $\kappa$ is a tuneable parameter which we typically set to $10^{3}$ in our experiments.
Using the $\bTheta_j$ from solving Eq.~\ref{eq:holonomy_opt} to rotate the tangent vectors after each tracing step, we largely eliminate artifacts caused by variations in curvature on the surface (Fig.~\ref{fig:holonomy_comp}, right).

One potential alternative is to instead smooth the outermost strip holonomies $R(S_{i,j})$ instead of wedge holonomies $R(\Delta_{i,j})$ to more directly smooth out the accelerations at the points on $\psi_j$; unfortunately, as we show in Appendix~\ref{app:holonomy}, this scheme produces undesirably large rotations.

\subsection{Map Interpolation}\label{sec:interp}
After tracing out the points, the next step is to interpolate these points to form a continuous map from the planar disc $D_R \subset \Tspace{\bp}$ ($R = nh$) to the surface: $q_\bp \colon D_R \to \M$.
Each input point $\bx = (x,y) \in D_R$ can be expressed in polar coordinates $(r, \theta)$, where $r \in [0,R]$, and from there, we evaluate the surface as a spline in polar coordinates.
First, we interpolate the isolines $\Psi$ using periodic cubic splines as functions of angle.
We then evaluate all these splines at $\theta$ and $\theta + \pi$ (i.e., the opposite angle) to get $2n+1$ unique points that sample two opposite radial curves connected at the map origin, which we can connect using a natural cubic spline $\tilde{\gamma}_{\theta}$ and evaluate at $r$.

\begin{figure}
  \includegraphics[width=\columnwidth]{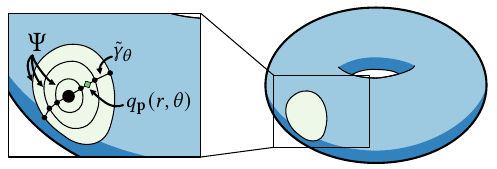}
  \caption{Using the same torus example from Fig.~\ref{fig:overview}, the map $q_\bp(r, \theta)$, $(r,\theta) \in \Tspace{\bp}$, is evaluated by first evaluating the isolines $\Psi$ at $\theta$ and $\theta + \pi$, interpolating those results to form a radial curve $\tilde{\gamma}_\theta$ passing through $\bp$, and evaluating this curve at $r$.}
  \label{fig:interp}
\end{figure}

\subsection{Logarithmic Map}\label{sec:logmap}
Along with the forward map $q_\bp$, it is also useful to have the inverse $q^{-1}_\bp$ (which we will call the logarithmic map even though $q_\bp$ is not exactly an exponential map).
Finding the exact inverse of the spline surface defined in Section~\ref{sec:interp} would require solving a non-linear optimization problem, so we instead mesh the spline by uniformly sampling several points in $D_R$, Delaunay triangulating those points, and pushing those points (and their connectivity) through $q_\bp$; with this mesh, we can evaluate $q^{-1}_\bp$ via closest point queries on the mesh with a small maximum radius (we use $10^{-2}$).
The point sampling can be improved by, e.g., drawing more samples from high-curvature portions of the map, but at a high enough resolution this basic scheme is sufficient for our applications.
For all examples in the paper, we sample 10000 points on the interior of $D_R$, and use $8m$ uniformly sampled points on $\partial D_R$ for the Delaunay triangulation boundary.
Optionally, to obtain a closer fit to the true surface, the spline-evaluated samples can be projected onto the zero isosurface, though this incurs an additional performance cost proportional to the number of samples, and we did not find it necessary for the results in the paper.

\section{Evaluation and Results}

\begin{table*}
  \centering
  \rowcolors{2}{white}{CornflowerBlue!25}
  \caption{
  Parameter listings of all the results in the paper, where $m$ is the number of radial curves, $n$ is the number of steps, and $h$ is the step size.
  The ``Size'' column lists the number of points in a point cloud, or the number of vertices in a mesh.
}
  \begin{tabular}{l r r r r r r r}
    \toprule
    \textit{Surface}  & \textit{Fig.}                             & \textit{Type}     & \textit{Size} & $m$ & $n$         & $h$                   & \textit{Smoothing Enabled?} \\
    Saddle            & \ref{fig:comparison}                      & Analytic Implicit & ---           & 50  & 10          & $1.0 \times 10^{-2}$  & Yes \\
    Armadillo         & \ref{fig:holonomy_regularization}         & Neural Implicit   & ---           & 50  & 20          & $5.0 \times 10^{-3}$  & Yes+No \\
    Cone              & \ref{fig:compositing}                     & SDF               & ---           & 50  & 12          & $1.0 \times 10^{-2}$  & No \\
    Bunny (coarse)    & \ref{fig:mesh_comparison} (top right)     & Mesh              & 3,485         & 20  & 20          & $2.5 \times 10^{-2}$  & Yes \\
    Bunny (fine)      & \ref{fig:mesh_comparison} (bottom right)  & Mesh              & 13,934        & 20  & 20          & $2.5 \times 10^{-2}$  & Yes \\
    Toy               & \ref{fig:teaser}                          & SDF               & ---           & 50  & 30          & $1.0 \times 10^{-2}$  & Yes \\
    Flute             & \ref{fig:teaser}                          & Analytic Implicit & ---           & 50  & 12          & $5.0 \times 10^{-3}$  & Yes \\
    Brain             & \ref{fig:teaser}                          & Mesh              & 73,820        & 50  & 10, 15, 20  & $5.0 \times 10^{-3}$  & Yes \\
    Owl               & \ref{fig:teaser}                          & Neural Implicit   & ---           & 50  & 5, 10       & $1.0 \times 10^{-2}$  & Yes \\
    Terrain           & \ref{fig:teaser}                          & Point Cloud       & 29,474,037    & 50  & 10          & $1.0 \times 10^{-2}$  & Yes \\
    Star Torus        & \ref{fig:curve}                           & SDF               & ---           & 50  & 40          & $1.0 \times 10^{-2}$  & Yes \\
    Shellfish         & \ref{fig:neural} (left)                   & Neural Implicit   & ---           & 50  & 15          & $1.0 \times 10^{-2}$  & Yes \\
    Einstein          & \ref{fig:neural} (middle)                 & Neural Implicit   & ---           & 50  & 20          & $5.0 \times 10^{-3}$  & Yes \\
    Metratron         & \ref{fig:neural} (right)                  & Neural Implicit   & ---           & 50  & 15          & $1.0 \times 10^{-2}$  & Yes \\
    Pothole           & \ref{fig:holonomy_comp}                   & SDF               & ---           & 50  & 20          & $1.0 \times 10^{-2}$  & Yes+No \\
    Spot              & \ref{fig:holonomy_breakdown}              & Mesh              & 2,930         & 50  & 20          & $1.0 \times 10^{-2}$  & Yes \\
    Shark             & \ref{fig:logmap_wraparound}               & Mesh              & 10,054        & 50  & 20          & $1.0 \times 10^{-2}$  & Yes \\
    \bottomrule
  \end{tabular}
  \label{tbl:params}
\end{table*}

\begin{table*}
  \centering
  \rowcolors{2}{white}{CornflowerBlue!25}
  \caption{Timing breakdown for maps generated on the surfaces in Fig.~\ref{fig:teaser}, with an in-depth breakdown for radial tracing. Times are given in \textit{milliseconds}. The root finding (``Root''), projected step (``Proj.''), and frame rotation (``Frame Rot.'') operations are done during each tracing substep, and smoothing is done after a full tracing step. Times are averaged over all maps on each surface, except for the brain and owl surfaces, where timings for the largest map are reported instead.}
  \begin{tabular}{l r r r r r r r r r r}
    \toprule
    \textit{Surface}  & \textit{Type}     & $n$ & \textit{Avg. \# Substeps} & \textit{Avg. Root}  & \textit{Avg. Proj}  & \textit{Avg. Frame Rot.}  & \textit{Avg. Smoothing} & \textit{Tracing}  & \textit{Fitting}  & \textit{Meshing} \\
    Toy               & SDF               & 30  & 2.000                     & 5.287               & 0.5509              & 0.5982                    & 0.6699                  & 414.3             & 2.336             & 196.0 \\
    Flute             & Analytic Implicit & 12  & 2.417                     & 10.85               & 2.219               & 1.304                     & 0.7731                  & 427.3             & 2.006             & 165.1 \\
    Brain             & Mesh              & 20  & 2.900                     & 14.66               & 0.8845              & 1.419                     & 0.6537                  & 502.5             & 2.005             & 158.3 \\
    Owl               & Neural Implicit   & 10  & 2.800                     & 14.75               & 1.089               & 1.294                     & 0.7760                  & 492.4             & 2.058             & 172.0 \\
    Terrain           & Point Cloud       & 10  & 2.733                     & 34.16               & 3.403               & 3.069                     & 0.7709                  & 1137              & 1.977             & 165.1 \\
    \bottomrule
  \end{tabular}
  \label{tbl:timings}
\end{table*}

\subsection{Implementation}
We implemented our code primarily in Python using the \textsc{PyTorch} library~\cite{Paszke2019} to have a framework for automatic differentiation, which we used to compute gradients of implicit functions and interface with trained neural implicits.
We used a simple feature grid architecture for the neural implicits used in the paper, based on the NGLOD architecture~\cite{Takikawa2021} with a single $128^3$ feature grid, and an MLP with two 128-unit hidden layers and softplus activations.
Point cloud implicits were obtained using a LogSumExp smooth distance function~\cite{Madan2022}.
Also, we used JIT compilation for analytic implicits and signed distance functions to speed up their evaluation.
To have comparable parameters across all experiments, we isotropically rescaled every object to fit within the cube $[-1,1]^3$~(Table~\ref{tbl:params}).

\subsection{Performance}
Asymptotically, each component of a single tracing step (projection, substepping, tridiagonal linear system solving) takes $O(m)$ time to advance the integration front from $\psi_j$ to $\psi_{j+1}$, so tracing $n$ steps is overall $O(mn$).
Fitting the surface also takes $O(mn)$ time to build all of the $n$ isoline splines, and $O(n)$ time to evaluate the surface spline at a point (mainly to build $\tilde{\gamma}_\theta$).
One key advantage of our approach is that the performance is \textit{output-sensitive}; in other words, it does not strongly depend on the complexity or size of the underlying surface (though in some cases like triangle mesh distance, each implicit distance query is $O(\log N)$ where $N$ is the number of mesh triangles).
For example, mesh-based heat methods~\cite{Sharp2019,Herholz2019} are slightly superlinear in $N$ from solving sparse matrices, but if $mn \ll N$, this is much worse than $O(mn)$.
Although the method of \citet{Herholz2019} provides fast local solves using only a small output-sensitive subset of vertex values, it still requires a global prefactorization that is superlinear and dependent on $N$~\cite{Herholz2017}.
In Table~\ref{tbl:timings}, we show a timing breakdown of the examples in Fig.~\ref{fig:teaser}, which are representative of the types of implicit functions we used throughout the paper.
All timings were recorded on a 2020 MacBook Pro with an M1 processor.
From these results, we see that none of the examples took more than 1.5s to complete, and the primary bottleneck is tracing, particularly root finding in each substep (line 8 of Alg.~\ref{alg:trace}).
Furthermore, by also looking at the parameters in Table~\ref{tbl:params} we observe that the terrain point cloud map took roughly twice as long to trace as the brain mesh map, despite the former having 400 times as many vertices as the latter, which confirms the output-sensitive nature of our algorithm.
Meshing the log map takes hundreds of milliseconds for every example, though this is because we evaluate the surface at over 10000 points to build the mesh, so this can easily be reduced if a coarser mesh is acceptable.
Furthermore, it is apparent that overall performance is proportional to the amount of time required to evaluate the implicit $f$ and its gradient.
Overall, these results are sufficient for interactive applications, even on the LiDAR terrain point cloud with over 29 million points, and suggest that speeding up the computation of $f$ and a faster root finding method can further improve performance.

\subsection{Parameter Analysis}

\begin{figure}
  \includegraphics[width=\columnwidth]{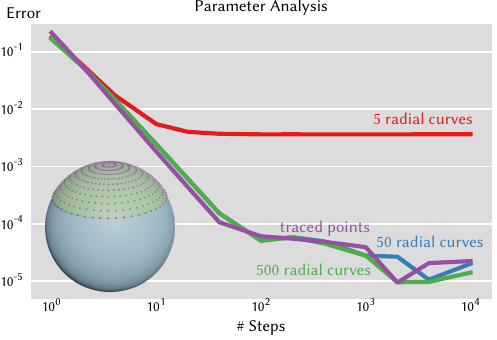}
  \caption{On an exponential map of the sphere, the mean error of our method compared to the analytical solution generally decreases as we increase the number of steps/decrease step size, but this eventually levels out for maps with more steps. More radial curves can reduce the error, but does not provide any additional benefit between 50 and 500 radial curves. The error along the traced radial points closely matches the overall map error.}
  \label{fig:error}
\end{figure}

The parameter $m$ (number of traced radial curves), $n$ (number of steps taken), and $h$ (step size) are the main quantities used to affect the accuracy of the parameterizations.
To demonstrate the effect of these parameters on a simple case with a known analytic solution, we ran our method on a unit sphere (with no substepping and no holonomy smoothing) using a range of values for each of these parameters (where $nh$, the total radius of the map in tangent space, is always $1$) and computed the average $L^2$ error of our output from the analytic exponential map, at 2000 samples inside the parameterized disc in tangent space and 500 points along the boundary of the disc~(Fig.~\ref{fig:error}).
Overall, we observe that both adding more steps (and correspondingly smaller step sizes) and more radial curves can improve the results, but after enough steps/radial curves, adding more steps/radial curves does not seem to improve the results.
In particular, increasing $m$ from 5 to 50 significantly reduces the error, though there is little difference between using 50 or 500 radial curves; as such, we use $m = 50$ for most of the results in the paper.
Furthermore, we see that the error levels out under refinement of $h$, but at different points when using 5 radial curves (10 steps) and 50/500 radial curves (100 steps).
To further investigate the source of the error, we also plotted the error from our traced points along a single radial curve for each step size (since this is on a sphere, there is little difference between each individual radial curve).
This error curve follows a very similar trend to the larger $m$ error curves which decrease and eventually level out around $10^{-5}$, which suggests that once there are sufficiently many traced radial curves, the primary source of error is the tracing rather than the spline interpolation.
Since the error in the traced points accumulates over more steps, there is a tradeoff between accumulating larger per-step error over fewer steps, and accumulating smaller per-step error over more steps, which causes the error to level out under refinement.
Despite the lack of error reduction beyond $10^{-5}$, this is quite small, and as demonstrated in the examples throughout the paper, does not cause practical issues.

\begin{figure*}
  \includegraphics[width=\textwidth]{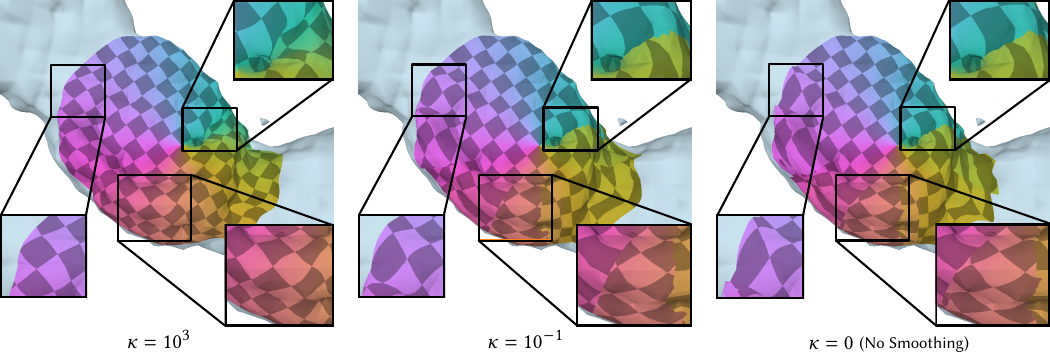}
  \caption{As we decrease $\kappa$, the effect of the smoothing is diminished. Near an indent in this neural implicit armadillo's arm (top right inset), decreasing $\kappa$ exchanges distortion near the indent (which diminishes away from the indent) for foldovers in the map. Furthermore, decreasing $\kappa$ can also affect regions farther away from the problematic indent, by inducing stretching (bottom left inset) and foldovers (bottom right inset) near the boundary of the map.}
  \label{fig:holonomy_regularization}
\end{figure*}

\begin{figure}
  \includegraphics[width=\columnwidth]{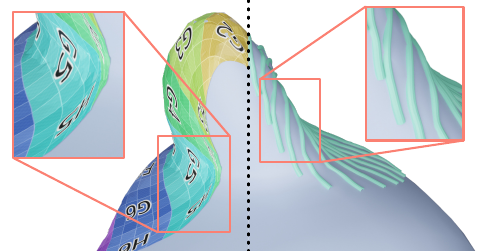}
  \caption{
    In the presence of particularly large surface features, holonomy smoothing produces large rotations and shears the integration front, causing curves to be near-parallel and overlap; the resulting map is also distorted. On this mesh of the log map, the symmetric Dirichlet energy~\cite{smith2015bijective} is 6.52, and the least squares conformal map (LSCM) energy~\cite{Levy2002} is 0.987, both higher than the averages shown in Fig.~\ref{fig:distortion_comparison}.
  }
  \label{fig:holonomy_breakdown}
\end{figure}

Holonomy smoothing strongly influences the output, and is controlled by a regularization parameter $\kappa$.
In Fig.~\ref{fig:holonomy_regularization}, we examine the effect of $\kappa$; as it decreases, the effect of the smoothing is diminished, and in the limit as $\kappa \to 0$, the smoothing is completely skipped.
We built a local parameterization for a neural implicit representing the Stanford armadillo, around a point near a small indent in the forearm created by isosurface noise.
With large $\kappa$ (i.e., strong smoothing), we see that the map has low distortion away from the origin, and although there is some distortion near the pothole itself, this distortion largely disappears as the map expands.
Meanwhile, as $\kappa$ decreases, the map's quality degrades; some regions fold over each other (seen through colormap discontinuities), while other regions significantly stretch away from the origin.
As a result, we exclusively use $\kappa = 10^{3}$ everywhere else in the paper.
However, in large map regions, holonomy smoothing can produce an undesirable shearing effect on the radial curves, where some curves end up becoming nearly parallel, an effect which primarily happens around large surface protrusions (Fig.~\ref{fig:holonomy_breakdown}).

Another parameter that can be controlled is the smoothing radius $\epsilon$, used for computing smoothed gradients $\widetilde{\nabla f}(\bx)$ in the ball $B_{\bx,\epsilon}$.
We used a Monte Carlo estimate of this smoothed gradient, by setting $\epsilon=10^{-4}$ and averaging 10 samples from $B_{\bx,\epsilon}$.
The variance of this estimate can affect the substepping procedure and produce different results across executions with the same input parameters, but we did not experience this in any of our experiments.
Nevertheless, we note that the accuracy of this estimate can be increased by adding more samples and/or reducing the smoothing radius.

\subsection{Comparisons}

\begin{figure*}
  \includegraphics[width=\textwidth]{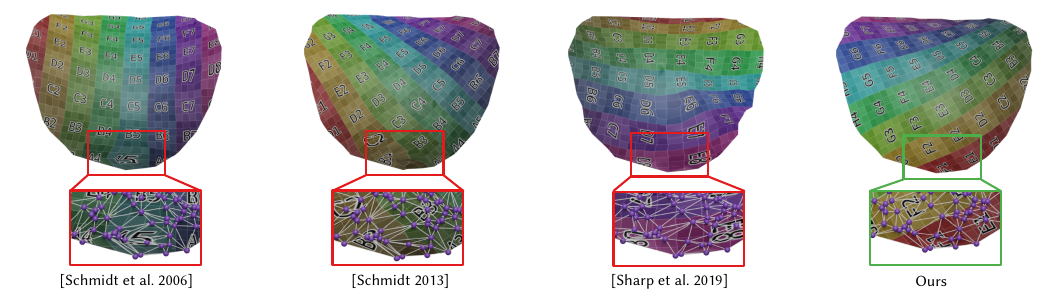}
  \caption{
  On a point set rejection sampled from a saddle surface, the original discrete exp map~\cite{Schmidt2006} (left) exhibits shearing artifacts away from the origin in the center. The more recent version~\cite{Schmidt2013} (middle left) with more accurate geodesics exhibits even more severe shearing. The vector heat method~\cite{Sharp2019} (middle right) is strongly influenced by the non-uniform point distribution and has distortion throughout the map. Our method (right) is the only one to exhibit low distortion everywhere, due to our lack of dependence on the sample points. The maps represent point sets connected via tangent space Delaunay triangulation, and the insets show the input samples and their Delaunay edges.
  }
  \label{fig:comparison}
\end{figure*}

Compared to prior local parameterization methods, our method can extract significantly more information out of implicit surfaces.
Rather than simply sampling the surface for points, and using nearest neighbor information to, e.g., build graphs or Laplacians, we generate the samples we need through tracing, while also being able to evaluate the generated map on other surface points.
However, our method does not provide precise control over the size of the resulting parameterization beyond integration parameters, which makes it difficult to directly compare to ``log map-based'' methods in prior work which require such information as input.
That said, our generated maps can still be evaluated on isolated surface samples after the maps have been expanded from a given surface origin, which forms the basis of our comparison methodology against prior work.
More specifically, to compare our method to prior work, we generated a local parameterization around a saddle point with 500 traced samples, and rejection sampled the same number of surface points from a 3-dimensional ball around the map origin in $\R^3$ such that all the points are contained in our map.
Then, we evaluated the log map of our method on those sample points, and also provided those points as input to both versions of discrete exponential maps (DEM)~\cite{Schmidt2006,Schmidt2013} and the vector heat method~\cite{Sharp2019}.
We used our own implementation for the DEM methods and the authors' released code for the vector heat method on point clouds\footnote{https://github.com/nmwsharp/potpourri3d}, which uses intrinsic Delaunay triangulations to build more numerically robust Laplacians.
For each map, we built a mesh connecting the samples by using the connectivity from the Delaunay triangulation computed in tangent space.
As seen in Fig.~\ref{fig:comparison}, all other methods produce high distortion away from the origin, and the vector heat method struggles throughout the entire map.
It is worth noting that these struggles are exacerbated by the na\"ive point sampling method, which tends to produce clusters of points that can be prevented by more sophisticated blue noise sampling techniques~\cite{Witkin1994,DeGoes2012}; nevertheless, our method produces low distortion throughout the map despite the sample quality, since we do not depend on them for generating the parameterization.

\begin{figure*}
  \includegraphics[width=\textwidth]{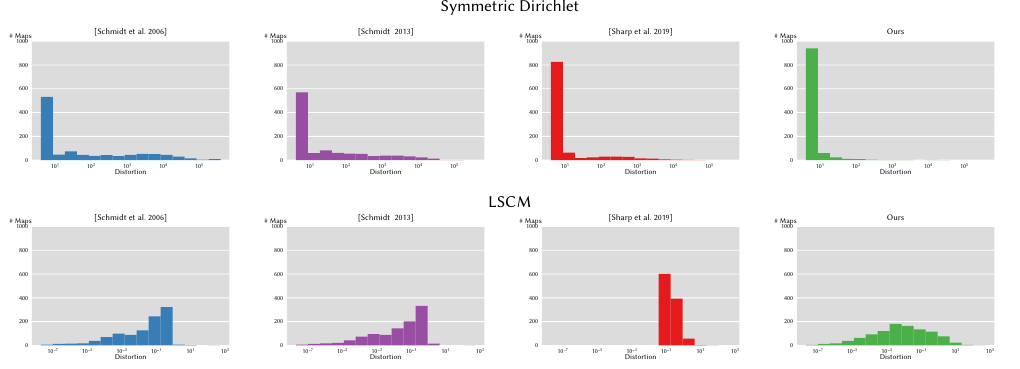}
  \caption{
  Across a dataset of surfaces, map positions, and map sizes, almost all of the maps produced by our method have near-minimum symmetric Dirichlet energy, whereas a significant fraction of the maps produced by other methods have higher energy. (top). Our method also produces maps with orders-of-magnitude lower median LSCM energy compared to other methods, though with more variance (bottom). All histograms are plotted on a log scale on the x-axis.
  }
  \label{fig:distortion_comparison}
\end{figure*}

We also used the same point surface point $\to$ log map setup to quantitatively investigate the distortion of our method's output compared to prior work.
More specifically, we used a dataset of over 100 surfaces collected by \citet{MPZ14}, and for each surface sampled 10 surface points as map origins, to obtain a set of over 1000 surface regions over which to build local parameterizations.
For half of these local surface regions, we sampled 500 surface points within a geodesic radius of 0.1 of the origin, and for the other half we sampled 1000 points within a geodesic radius of 0.2.
These points were then evaluated through the log map of each method used in Fig.~\ref{fig:comparison} and triangulated in tangent space to obtain connectivity for the sample points in $\R^3$.
We used the area-weighted average symmetric Dirichlet energy~\cite{smith2015bijective} and least-squares conformal map (LSCM) energy~\cite{Levy2002} to measure the degree to which the resulting parameterized meshes represent isometries and conformal maps, respectively.

The results of this experiment are provided as histograms in Fig.~\ref{fig:distortion_comparison}.
Across all methods, the symmetric Dirichlet distributions look quite similar, with the majority of maps exhibiting low distortion near the global minimum of 4, and a tail of maps with higher distortion.
However, our method (right) produces the largest fraction of low-distortion maps compared to the other tested methods, with over 900 of the maps contained in the lowest histogram bin, and as a result has a much ``thinner'' tail than the other methods.
The DEM methods in particular (left, middle left) have nearly half of their outputs outside of the lowest bin.
On the other hand, the LSCM distributions vary significantly across the different methods: DEM methods exhibit a mode near the right side of the distribution near 1, the vector heat method (middle right) has a much narrower distribution but with a lower mode near $10^{-1}$, and our method has a much wider variance but a mode of $10^{-3}$, orders of magnitude lower than the other methods.
We can see one such example of a high LSCM energy in Fig.~\ref{fig:holonomy_breakdown}, where the map has an LSCM energy of nearly 1.
Overall, we see that the combination of geodesic tracing, substepping, and holonomy smoothing produces near-isometric and near-conformal maps in most cases, which quantitatively affirms Fig.~\ref{fig:comparison}, but can produce non-conformal results when holonomy smoothing shears regions with large variations in curvature, as seen in Fig.~\ref{fig:holonomy_breakdown}.

\begin{figure*}
  \includegraphics[width=\textwidth]{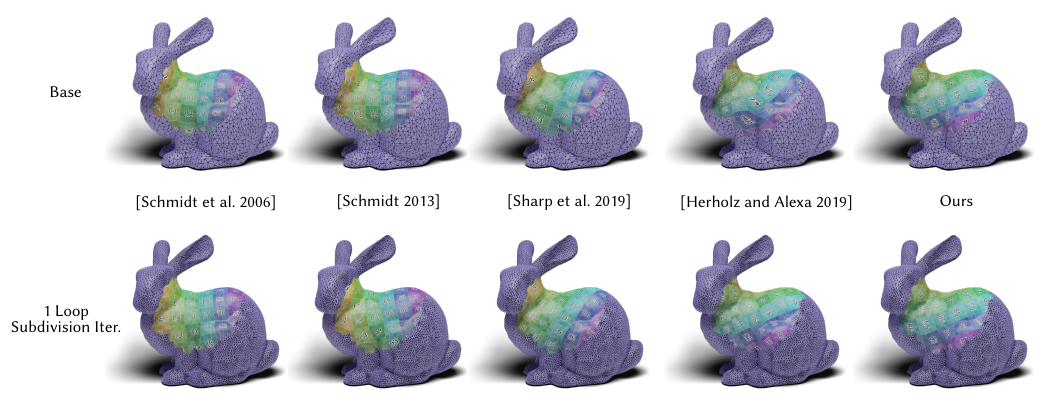}
  \caption{
  On a coarse bunny mesh (top), a local parameterization generated by our method (right) avoids the severe artifacts near the boundary of the map (left, centre left) and angular distortion near the origin (centre, centre right) seen by other methods. All methods improve after one level of Loop subdivision (bottom), but the DEM-based methods still possess boundary artifacts (left, centre left). Please zoom in to see map details.
}
  \label{fig:mesh_comparison}
\end{figure*}

Since we trace points along the surface, another interesting property of our method on triangle meshes in particular is that our result quality is independent of the underlying triangulation.
In Fig.~\ref{fig:mesh_comparison} we compare our method to both DEM methods, the vector heat method, and another heat-based method~\cite{Herholz2019}, for which we used our own implementation, taking care to use intrinsic Delaunay operators from \textsc{libigl}~\cite{libigl} (though our implementation does not use the accelerated solve, since a public implementation of the specialized prefactorization data structure is unavailable). 
For this experiment, we computed the 10-ring of an origin vertex (i.e., all vertices at most 10 edges away from the origin) on a coarse Stanford bunny mesh with 3500 vertices, and used these as the evaluation points for all methods.
The DEM methods only used these points and their derived connectivity from the mesh, while we used the entire mesh to define the required operators for the heat methods (since we would change the boundary conditions if we only used the local patch).
Our method does not build a map from pre-selected discrete points (aside from the map origin), so to make a best-effort comparison we ran our method so that the resulting map both contained the same 10-ring and used a comparable point budget (400 traced samples vs. 453 vertices in the 10-ring --- see Table~\ref{tbl:params} for the parameters we used).
The original mesh connectivity was used to connect the 10-ring evaluation points into a continuous patch for all methods.
We see that our method produces by far the least distortion in the map, and just like in Fig.~\ref{fig:comparison}, the DEM methods exhibit distortion far away from the origin, while the heat-based methods struggle throughout the map, since the triangulation is not of a high enough quality to produce well-conditioned operators.
After one level of Loop subdivision (which produces a mesh with roughly 14000 vertices), we ran the same experiment on the 20-ring to cover the same region on the surface (now containing 1720 vertices), and did not change the parameters in our method (though we should clarify that the underlying surface is now the \textit{subdivided} mesh and not the original mesh).
Now we see that our method produces virtually the same result as before, the heat-based methods perform much better with very little distortion, and the DEM methods improve but still exhibit artifacts at the boundary of the map.
However, these other methods only achieve this quality with 4 times the number of points that we need to achieve similar quality, and cannot achieve this at all on a coarser mesh.

\subsection{Applications}

\subsubsection{Local Texturing}
A variety of examples of texturing different representation types is provided in Fig.~\ref{fig:teaser}.
We show several decals on each surface, where each decal is represented by a meshed version of the spline surface, and can see that none of the resulting images are distorted, while still conforming to the original surface.
The terrain point cloud, owl neural implicit, and brain mesh all have challenging geometry with surface bumps, ridges, and divets that make parameterization more difficult.
By placing maps in small regions, however, even regions like the owl's plumage and the heavily folded surface of the brain can be textured.

\begin{figure}
  \includegraphics[width=\columnwidth]{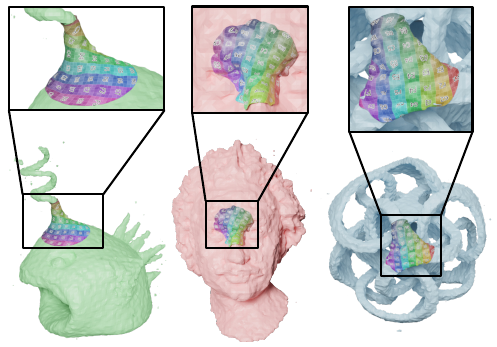}
  \caption{
  We obtain high-quality maps on a variety of neural implicit surfaces (rendered as meshes reconstructed with marching cubes), even in the presence of noisy reconstruction artifacts and fine ground truth details. The local maps reveal surface bumps and folds that the marching cubes-reconstructed surface mesh does not capture at its configured resolution.
}
  \label{fig:neural}
\end{figure}

To further illustrate map quality on highly detailed surfaces and high-genus surfaces, we show some more maps on neural implicits in Fig.~\ref{fig:neural}.
All three surfaces are bumpy from reconstruction error, but due to holonomy smoothing, our method is largely able to ignore the effects of both reconstruction noise and ground truth high-frequency features (e.g., the bridge of the Einstein bust's nose) on the radial path trajectories, and produce low-distortion maps that still conform to the underlying surfaces.

\subsubsection{Multi-Valued Logarithmic Maps}

\begin{figure}
  \includegraphics[width=\columnwidth]{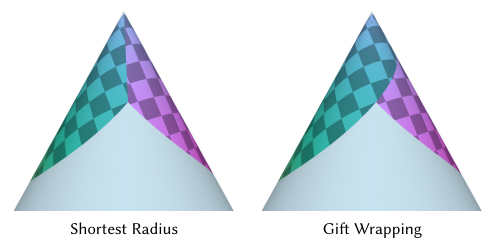}
  \caption{Although local geodesics can fold over, by computing the map from tangent space rather than from the embedding space, we can obtain a multi-valued log map, which we can use to find the globally shortest paths (left) and ``gift wrap'' the map around a cone tip (right).}
  \label{fig:compositing}
\end{figure}

Unlike other methods, which compute maps by assigning texture coordinates to surface points, our method operates in the opposite direction and instead traces out surface paths corresponding to radial lines in tangent space.
This allows us to support multi-valued logarithmic maps as well, which can encode all the local geodesics that pass through a point on the surface (and smoothing could be disabled in such cases).
As a proof of concept, we show some preliminary visualization results that take advantage of this property, by compositing multi-valued log map values in a sphere tracer.
Compositing is done by taking all ray intersections with the log map and sorting the intersections (near the closest ray intersection to the underlying surface) based on user specification.
The shortest radius mode reproduces global geodesics by taking the point with the smallest radius (reproducing the seam we would expect with global geodesics).
Alternatively, we can ``gift wrap'' the map by taking the value with the smallest phase, which, to our knowledge, is not an effect that can be produced by other methods~(Fig.~\ref{fig:compositing}).

\subsubsection{Curve Drawing on Surfaces}
\begin{figure}
  \includegraphics[width=\columnwidth]{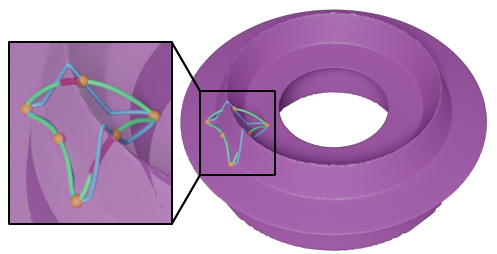}
  \caption{Using local parameterizations, we can draw $\kappa$-curves directly on an isosurface (blue) by placing constraint points on the surface (orange) and optimizing control points in tangent coordinates. Solving in the ambient space with the same constraint points leads to intersections with the surface (green). The inset shows a translucent view of the curves, where the regions of the ambient curve that cut inside the surface are shown in red.}
  \label{fig:curve}
\end{figure}

It is also useful to be able to ``connect'' nearby parameterizations by composing the inverse of one with the forward evaluation of another (i.e., $q_{\bp_2} \circ q^{-1}_{\bp_1}$), to convert surface points between different local coordinate systems, which we illustrate by lifting the $2D$ $\kappa$-curves~\cite{Yan2017} algorithm onto surfaces using our method.
With local parameterizations we can solve for the curve's control points in tangent coordinates, and evaluate each curve segment in a separate local map, enforcing that all points lie on the surface by construction.
For details on how this is done, please see Appendix~\ref{app:curves}.
We show our results in Fig.~\ref{fig:curve}; the curve produced with our method exactly conforms to the surface, while attempting the same process directly in $\R^3$ results in the curve cutting through the surface.

\section{Limitations, Future Work and Conclusion}

\begin{figure}
  \includegraphics[width=\columnwidth]{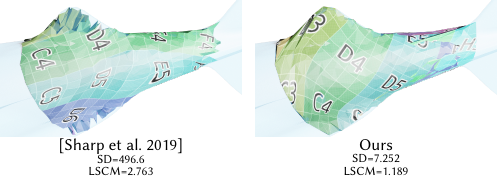}
  \caption{
    A challenging example from the dataset of \citet{MPZ14} where our method's log map triangulated from surface samples exhibits high distortion (right), with shearing near the shark's dorsal fin, and wrapping around itself near the tail (unlike Fig.~\ref{fig:compositing} which used a custom renderer, the Blender renderer used here does not support map-aware compositing). However, when the vector heat method~\cite{Sharp2019} (left) is applied to the same set of sample points, it also struggles to reproduce the dorsal fin, though it produces an injective map. The two distortion metrics from Fig.~\ref{fig:distortion_comparison} are given as well, where ``SD'' represents symmetric Dirichlet energy, and ``LSCM'' represents LSCM energy.
  }
  \label{fig:logmap_wraparound}
\end{figure}

The primary technical limitation of our method is that our maps are restricted to discs in tangent space, which also restricts their maximum radii.
Unlike ``log map-based'' methods on meshes~\cite{Sharp2019,Herholz2019}, which assign tangent coordinates to each vertex and indirectly create a star around the origin in tangent space, our method restricts the geometry of the domain and thus cannot expand into large maps across complex surfaces without intersecting or rapidly separating (adjacent) geodesics, which lead to a poorly interpolated spline surface.
Even holonomy smoothing cannot entirely ameliorate this issue, particularly around large protrusions on the surface, because it will try to emulate a surface of constant Gaussian curvature that will eventually cause adjacent geodesics to run nearly parallel with each other (Fig.~\ref{fig:holonomy_breakdown}).
However, even prior work such as the vector heat method struggles in these challenging scenarios; a challenging example from Fig.~\ref{fig:distortion_comparison} is shown in Fig.~\ref{fig:logmap_wraparound} where both our method and the vector heat method produce distorted results.
The parallel transport assumption along the current integration front $\psi_j$ used in holonomy smoothing can also degrade in accuracy as the geodesics separate over several tracing iterations, though this can be remedied by tracing more geodesics to reduce the inter-geodesic separation distances.
Thus, a method that can separate varying angular density between radial geodesics, varying termination steps for each geodesic, and an interpolation method that can accurately stitch such geodesics together, is required for a fully robust ``exp map-based'' solution to the local parameterization problem, though our much simpler approach already produces excellent results on many complex surfaces and previously unsupported geometric representations.

There are many more applications of local parameterizations beyond the ones we showed in this paper.
For example, conducting simulation algorithms on surfaces, such as fluid simulation, particle dynamics, and even algorithms for more esoteric phenomena like ice crystal growth~\cite{Kim2003}, are a few examples that can be more easily unlocked by high-quality local parameterizations.
Another application is in running generative models such as diffusion models~\cite{Song2020} and normalizing flows~\cite{Kobyzev2020} on surfaces.
Although there has been some work on generalizing the latter to surfaces~\cite{Lou2020}, it is restricted to surfaces where analytic exponential maps are known, which greatly limits its applicability.
More broadly in computer graphics, surface-based algorithms are often relegated to cases where the surface is explicitly provided, such as with triangle meshes, but with the emergence of new neural geometric representations like neural implicits, neural radiance fields~\cite{Mildenhall2021}, and Gaussian splats~\cite{Kerbl2023}, their adoption in wider applications is limited by the need to design tailored algorithms for each representation.
We therefore believe that general-purpose algorithms like ours are an important step towards making more geometry widely usable.

\begin{acks}
This project was funded in part by an NSERC Discovery Grant (RGPIN-2023-05120) and an Ontario Early Researchers Award.
The first author was funded by an NSERC Canada Graduate Scholarship --- Doctoral.
We thank Victor Rong, Honglin Chen, Silvia Sell\'an, and Towaki Takikawa for proofreading.
\end{acks}

\bibliographystyle{ACM-Reference-Format}
\bibliography{implicit-exp-maps}

\appendix

\section{Instability of Strip Holonomy}\label{app:holonomy}

Here we show that smoothing the holonomy of the outermost strips $S_{i,j}$ is unstable, or more precisely, we show that strip holonomy smoothing produces undesirably large rotation angles.
The strip holonomy is given by
\begin{equation}
  \label{eq:strip_holonomy_app}
  R(S_{i,j}) = (\theta_{i,j} + \phi_{i,j} - \theta_{i+1,j}) - (\theta_{i,j-1} + \phi_{i,j-1} - \theta_{i+1,j-1}),
\end{equation}
and the smoothing optimization is
\begin{equation}
  \label{eq:holonomy_opt_app}
  \bTheta^*_j = \min_{\bTheta_j} \sum_i R(S_{i,j})^2 + \frac{1}{\kappa} \| \bTheta_j \|^2.
\end{equation}
The solution to Eq.~\ref{eq:holonomy_opt_app} is given by the linear system
\begin{equation}
  \label{eq:holonomy_soln_app}
  \left( \Lap + \frac{1}{\kappa} \Id \right) \bTheta_j = \bPhi_j - \bPhi_{j-1} - \Lap \bTheta_{j-1},
\end{equation}
where $\Lap$ is the 1D (positive-semidefinite) Laplacian, $\Id$ is the identity matrix, and $\bPhi_j$ is defined coordinate-wise by $(\bPhi_j)_i = (\phi_{i-1,j} - \phi_{i,j})$.
However, we typically want $\frac{1}{\kappa}$ to be small for smoothing to be effective, so we will drop the $\frac{1}{\kappa} \Id$ term.
Since $\bPhi_0 = \bTheta_0 = \mathbf{0}$, we have $\Lap \bTheta_1 = \bPhi_1$, and then $\Lap \bTheta_2 = \bPhi_2 - 2\bPhi_1$, $\Lap \bTheta_3 = \bPhi_3 - (2\bPhi_2 - 2\bPhi_1)$, and in general
\begin{equation}
  \label{eq:holonomy_soln_full_app}
  \Lap \bTheta_j = \bPhi_j - 2\sum_{k=1}^{j-1} (-1)^{j-k-1}\bPhi_k.
\end{equation}
Since $\Lap^{-1}$ has eigenvalues $\gg 1$, and the upper bound of the right-hand side of Eq.~\ref{eq:holonomy_soln_full_app}, $\| \bPhi_j \| + 2\sum_{k=1}^{j-1} \| \bPhi_k \|$, grows in magnitude as $j$ increases, the upper bound of $\| \bTheta_j \|$ also grows in magnitude as $j$ increases.
Although this is not a tight bound, from Eq.~\ref{eq:holonomy}, we can interpret $\bPhi_j$ as the difference in wedge holonomy between adjacent $\Delta_{i,j}$ before smoothing, and since the total (absolute) curvature contained in $\bigcup_i \Delta_{i,j}$ increases as $j$ increases, we can expect adjacent wedge holonomies to diverge and the norm of holonomy differences to increase with $j$, with the exception of constant curvature surfaces where $(\bPhi_j)_i = 0$.
As such, $\| \bTheta_j \|$ generally increases with $j$, which matches what we observed in practice, and thus strip holonomy smoothing is unsuitable for a procedure that is primarily intended to make small adjustments to radial paths.

In contrast, the solution to wedge holonomy smoothing in Eq.~\ref{eq:holonomy_opt} drops the alternating summation:
\begin{equation}
  \label{eq:wedge_holonomy_soln}
  \left( \Lap + \frac{1}{\kappa}\Id \right) \bTheta_j = \bPhi_j.
\end{equation}
As such, its behaviour is much more benign and only depends on $\bPhi_j$ rather than an alternating sum of previous $\bPhi_k$.

\section{Curve Drawing on Surfaces}\label{app:curves}

To elaborate on our $\kappa$-curves implementation on surfaces, we replaced the global step of the $\kappa$-curves local-global solver with 10 iterations of a Gauss-Seidel solver, where every iteration solves for the $c_1$ control points in sequence (using the notation from \citet{Yan2017}).
Each $c_1$ is associated with a local map and its corresponding origin $\bp_k$, so they are initialized to be $\mathbf{0}$ in $\Tspace{\bp_k}$ at the start of the solve.
For the solve to work, each $c_1$ must also be covered by the adjacent local maps along the curve, and stay within those maps throughout the solve.
Although we do not try to ensure that the optimization stays within the parameterized tangent disc for each map, we find that in practice we never encountered solver issues if an interactive user is careful to place each high-curvature constraint point (the $\bp_k$) in regions that overlap the previous point in the sequence, as well as the very first point when closing the curve.

\end{document}